\documentclass{aa}

\newcommand{\chandra}{{\it Chandra}}

\newcommand{\xmm}{{\it XMM-Newton}}

\newcommand{\kepler}{{\it Kepler}}

\begin{document}

\title{Searching for the 1\,mHz variability in the flickering of V4743\,Sgr: a Cataclysmic Variable accreting at a high rate}
\titlerunning{flickering in V4743\,Sgr}
\authorrunning{Dobrotka et al.}

\author{A.~Dobrotka \inst {1}, M.~Orio \inst {2,3}, D.~Benka \inst {1} and A.~Vanderburg \inst {2}}

\offprints{A.~Dobrotka, \email{andrej.dobrotka@stuba.sk}}

\institute{Advanced Technologies Research Institute, Faculty of Materials Science and Technology in Trnava, Slovak University of Technology in Bratislava, Bottova 25, 917 24 Trnava, Slovakia
\and
			Department of Astronomy, University of Wisconsin 475 N. Charter Str. Madison, WI 53706
\and
			INAF - Astronomical Observatory Padova, vicolo dell'Osservatorio 5, 35122 Padova, Italy
}

\date{Received / Accepted}

\abstract
{}
{A few well studied cataclysmic variables (CVs) have shown discrete characteristic frequencies of fast variability; the most prominent ones are around log($f$/Hz) $\simeq$ -3. Because we still have only small number statistics, we obtained a new observation to test whether this is a general characteristic of CVs, especially if mass transfer occurs at a high rate typical for dwarf nova in outbursts, in the so called ``high state''.}
{We analyzed optical \kepler\ data of the quiescent nova and intermediate polar V4743\,Sgr. This system hosts a white dwarf accreting through a disk in the high state. We calculated the power density spectra, and searched for break or characteristic frequencies. Our goal is to assess whether the mHz frequency of the flickering is a general characteristic.} 
{V4743\,Sgr has a clear break frequency at log($f$/Hz) $\simeq$ -3. This detection increases the probability that the mHz characteristic frequency is a general feature of CVs in the high state, from 69\% to 91\%. Furthermore, we propose the possibility that the variability is generated by similar mechanism as in the nova-like system MV\,Lyr, which would make V4743\,Sgr unique.}
{}


\maketitle

\section{Introduction}
\label{introduction}

Cataclysmic variables (CVs) are interacting binaries powered by an accretion process. The mass is transferred from a main sequence companion star and falls onto the white dwarf (WD). In the absence of a strong magnetic field, an accretion disc forms. CVs are divided into subclasses based on characteristic variability features. Dwarf novae (DN) show quasi-regular outbursts lasting several days, repeated on  time scales of 10 - 100 days (see \citealt{warner1995} for review). Nova-like binaries, especially the  VY\,Scl systems, spend most of their lifetime in a ``high state'' in which the accretion rate $\dot{m}_{\rm acc}$ is so high that DN outbursts are suppressed. The exact value of $\dot{m}_{\rm acc}$ depends on the rest of the physical parameters of the system, but is generally around a few 10$^{-10}$ M$_\odot$ yr$^{-1}$. Most classical novae in the years immediately after an outburst may also be in a stage of enhanced mass transfer, while they may be ``hibernating'' for most of the period between outbursts (\citealt{hillman2020}). Both classes of objects, VY Scl and novae,  seem to have  stable accretion disks and are good targets to study the accretion process.

The disc instability model explains the dwarf nova cycle (see \citealt{lasota2001} for a review). The DN outbursts are caused by a viscous-thermal instability triggered by changes in the ionization state of hydrogen while matter flows through the disc (\citealt{hoshi1979}, \citealt{meyer1981}, \citealt{lasota2001}). Thus there are two basic states, recurring in a limit cycle: the high state (DN outbursts) when $\dot{m}_{\rm acc}$ through the disc increases, and the low state with lower $\dot{m}_{\rm acc}$ (quiescence; see \citealt{osaki1974} for the original idea). The accretion disc is fully developed up to the WD in the high state, while it is truncated in the low one. The high state is optically brighter than the low state.

The fast stochastic variability, which we call flickering, is a typical manifestation of the underlying accretion process. The three most important observational characteristics of the flickering are; 1) linear correlation between variability amplitude and log-normally distributed flux (so called rms-flux relation) observed in all accreting systems (see \citealt{scaringi2012b}, \citealt{vandesande2015} for CVs), 2) power density spectra (PDS) in the shape of a red noise or band limited noise with characteristic frequencies (\citealt{scaringi2012a}, \citealt{dobrotka2014}, \citealt{dobrotka2016} describe how this occurs in CVs) and 3) time lags, in which the flares reach their maxima slightly earlier in the blue than in the red (\citealt{scaringi2013}, \citealt{bruch2015}).

In the light curves of several CVs, characteristic frequencies have been identified in the PDSs. Ground observations in the optical range have usually shown a single break frequency like in UU\,Aqr (\citealt{baptista2008}) and KR\,Aur (\citealt{kato2002}). Detecting multiple PDS components requires light curves obtained in long exposures. These were provided by the \kepler\ satellite for MV\,Lyr (\citealt{scaringi2012a}), V1504\,Cyg (\citealt{dobrotka2015}) and V344\,Lyr (\citealt{dobrotka2016}). Additional observations with \xmm\ in the X-ray range have allowed detection of the same frequencies detected in optical in MV\,Lyr (\citealt{dobrotka2017}) and possibly in V1504\,Cyg (Dobrotka et al., in preparation). Other systems like VW\,Hyi, WW\,Cet and T\,Leo were observed in X-rays, revealing a single characteristic frequency (\citealt{balman2012}) or multiple components in SS\,Cyg (\citealt{balman2012}) or RU\,Peg (\citealt{dobrotka2014}).

\citet{dobrotka2020} summarized all these detections, suggesting that two characteristic frequencies can exist in the low state, with an additional and most prominent frequency with log($f$/Hz) $\simeq$ -3 in the high state. However, this suggestion was based on only 12 detections. The frequency distribution can be easily explained by  the random appearance of a uniform distribution. More detections are needed to construct a significant histogram, which is the aim of this paper.

The flickering phenomenon and its frequencies, in fact, bear direct evidence of the accretion mode. \citet{scaringi2014} interpreted the main feature at log($f$/Hz) $\simeq$ -3 in the PDS calculated from \kepler\ data of MV\,Lyr as due to a hot, geometrically thick disc (hot X-ray corona), surrounding a cool, geometrically thin disc. The propagating mass accretion fluctuations (\citealt{lyubarskii1997}, \citealt{kotov2001}, \citealt{arevalo2006}) in the corona generate X-ray variability with characteristic frequency of log($f$/Hz) $\simeq$ -3. The X-rays are reprocessed by the geometrically thin disc into optical variability. 

\citet{dobrotka2017} further elaborated this model on the basis of observations, proposing that two different regions in a CV may emit X-rays, not just the boundary layer as it is usually assumed. One region is the standard boundary layer between the geometrically thin disc and the WD, the second must be the boundary between the geometrically thick corona and the WD, or the corona itself.

However, the ratio of X-ray to optical luminosity is of the order of 0.1 (see e.g. \citealt{dobrotka2020}), which is too low to explain the observed optical and UV variability. \citet{dobrotka2019} studied the shot profile of the flickering in the \kepler\ data of MV\,Lyr, finding high and low amplitude components with frequencies close to log($f$/Hz) $\simeq$ -3. The small amplitude one can be indeed explained with the reprocessing scenario (see also \citealt{dobrotka2020}). The variability is attributed to two separate sources: the geometrically thin disc, and the reprocessed X-rays of the geometrically thick corona. 

\section{V4743\,Sgr and the new observations}
\label{v4743sgr}

In this study, we analyse the variability of a CV system that does
 not show DN outbursts, and for several reasons is very likely to be found in a high state (that is, with a hot and ionized disk); the recent nova V4743\,Sgr (Nova Sgr 2002c).

V4743\,Sgr was discovered in outburst in September 2002 by \cite{haseda2002}. While it was still in outburst and had turned into a luminous supersoft X-ray source, \citet{ness2003} found variability with a 0.75\,mHz frequency in \chandra\ data taken 180 days after optical maximum. More detailed analyses of these \chandra\ data and of \xmm\ X-ray data taken 196 days after maximum were later presented by \cite{leibowitz2006} and \citet{dobrotka2010}. The 0.75\,mHz variation was also still present in X-rays even at quiescence, as late as day 1286 after maximum. However, this dominant feature was double peaked in nature in the observations taken 180 and 196 days after the outburst.

\cite{kang2006} presented ground optical photometry and detected two periods of 6.7\,hr and $\sim 24$\,min. The authors attributed the lower frequency signal to the orbital period and, assuming that the 0.75\,mHz signal (22\,min) present in X-rays is the rotation period of the central WD, the optical period of 24\,min was interpreted as the beat period between the orbital and rotation period of the WD, respectively. This interpretation was confirmed by \citet{zemko2018} by analysing the \kepler\ light curve.

Detection of the rotation period in X-rays suggests an intermediate polar (IP) nature of V4743\,Sgr. \citet{zemko2016} analyzed the X-ray observations they had taken at quiescence. The X-ray spectra did have characteristics in common with known IPs, with  thermal plasma and a partially covering absorber, and a supersoft blackbody-like component, possibly originating from the polar regions irradiated by an accretion column.

The enigmatic double peak nature of the main feature at 0.75\,mHz was resolved by \citet{dobrotka2017b}. Using simulations, these authors showed that the double peak is produced by a single frequency of 0.75\,mHz with variable amplitude.

V4743\,Sgr has not shown DN outbursts during \kepler\ monitoring (\citealt{zemko2018}) and is likely to be in the enhanced mass transfer rate predicted by the hibernation scenario, with the disc of V4743\,Sgr in a hot and ionized high state. The binary was in Field 7 of the \kepler\ telescope (\citealt{borucki2010}) during the K2 mission between 2015 Oct 04 and 2015 Dec 26, so we requested to obtain data (EPIC 216631947) with the short cadence in which the spacecraft saved and downloaded images of the target every 58.85 seconds, as opposed to Kepler's usual 29.4 minute ``long cadence'' sampling rate.

\begin{figure}
\resizebox{\hsize}{!}{\includegraphics[angle=-90]{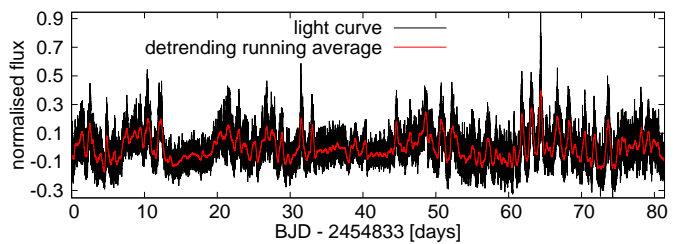}}
\caption{\kepler\ light curve of V4743\,Sgr. The flux has been normalised by dividing by the median flux of the observation. The time is in standard \kepler\ time.}
\label{lc_v4743sgr}
\end{figure}

The \kepler\ light curve is shown in Fig.~\ref{lc_v4743sgr}. Before the flux normalisation, systematic corrections were applied\footnote{The \kepler\ team estimates the contribution of scattered background light and subtracts that off. The background is a large source of photon noise, so in some cases with faint stars, where the background is much larger than the flux from the star, the estimated brightness may be negative in some cases.}. The data were barycentre corrected.

While the ``long cadence'' data allowed detection of the beat period between the orbital and the rotation period (\citealt{zemko2018}), with the ``short cadence'' we wanted to study in detail the flickering feature of this recent nova returned in quiescence, which is also an IP. An interesting question is whether IPs, with their truncated disks, satisfy disc conditions for the scenario described above and whether we can expect to detect similar characteristic frequencies of flickering.

The truncation of the inner disc is generated by a magnetic pressure balancing the ram pressure of the accretion flow. This defines a magnetospheric radius
\begin{equation}
r_{\rm m} = 9.8 \times 10^8 \left( \frac{\dot{m}_{\rm acc}}{10^{15}\,{\rm g}\,{\rm s}^{-1}} \right)^{-2/7} m_{\rm 1} \left( \frac{\mu}{10^{30}\,{\rm G}\,{\rm cm}^3} \right)^{4/7}\,{\rm cm},
\label{equation_rm}
\end{equation}
where $m_{\rm 1}$ is the WD mass and $\mu$ is the magnetic moment of the WD. If the mass transfer rate in post-novae is enhanced, this enhances the mass accretion rate $\dot{m}_{\rm acc}$. When $\dot{m}_{\rm acc}$ is larger, $r_{\rm m}$ is smaller. Equation (\ref{equation_rm}) describes how the inner disc truncation at $r_{\rm m}$ decreases if the $\dot{m}_{\rm acc}$ parameter rises keeping all other values like $m_{\rm 1}$ and $\mu$ fixed.

However, V4743\,Sgr is quite a special case. If $\dot{m}_{\rm acc}$ in post-novae is enhanced, it can be 3 order of magnitudes ($10^{-8}$ vs. $10^{-11}$\,M$_{\rm \odot}$\,yr$^{-1}$) larger than in the average IP in which the disc truncation is large. The decrease of the truncation radius $r_{\rm m}$ can be large. Based on Equation (\ref{equation_rm}) it can reach 14\% of the value with quiescent $\dot{m}_{\rm acc}$. It is difficult to establish whether this makes the accretion disc similar to non-magnetic systems, but we suggest that V4743\,Sgr is unlikely to be an IP with a large inner disc radius.

\section{PDS analysis}

\subsection{Method}

V4743\,Sgr was observed for a long time, allowing a more detailed time evolution study, therefore we studied separately each consecutive 10-day time interval. For the PDS study we divided each 10 day light curve into 10 subsamples, and using the Lomb-Scargle algorithm (\citealt{scargle1982}) normalised by the total variance (\citealt{horne1986}) we calculated log-log periodogram for each of these subsamples. We divided these periodograms into equally spaced bins of 0.05\,dex, and averaged all periodogram values within these bins in order to get {the} mean power $p$. The only condition was to include at least 20 periodogram points per bin in order to have a Gaussian distribution. The averaging is performed over log($p$) rather than $p$ following \citet{papadakis1993}, and such averaging yields symmetric errors (see e.g. \citealt{vanderklis1989}, \citealt{aranzana2018}). The PDS low frequency end is defined by the duration of each light curve subsample while the high frequency end is set by the Nyquist frequency.

\subsection{Results}

We downloaded the short-cadence pixel data from the Mikulski Archive for Space Telescopes (MAST\footnote{\url{https://archive.stsci.edu/k2/search_retrieve.html}}). We extracted the light curve and removed systematics due to the spacecraft's drift\footnote{The drift is generated by inability of the spacecraft to maintain precise pointing due to reaction wheels failure. The telescope re-points itself on timescales of 6 hours.} following \citet{vanderburg2014} and \citet{vanderburg2016}.

The light curve in Fig.~\ref{lc_v4743sgr} shows obvious variability on a time scale of about a day. Flare-like phenomena follow one another or are separated by constant flux intervals. In order to analyse equivalent data subsamples (with the same long-term trend) we detrended the light curve with a running average using 401 points in the running window\footnote{200 points to the left, 200 points to the right and the detrended point itself.}.

\begin{figure}
\resizebox{\hsize}{!}{\includegraphics[angle=-90]{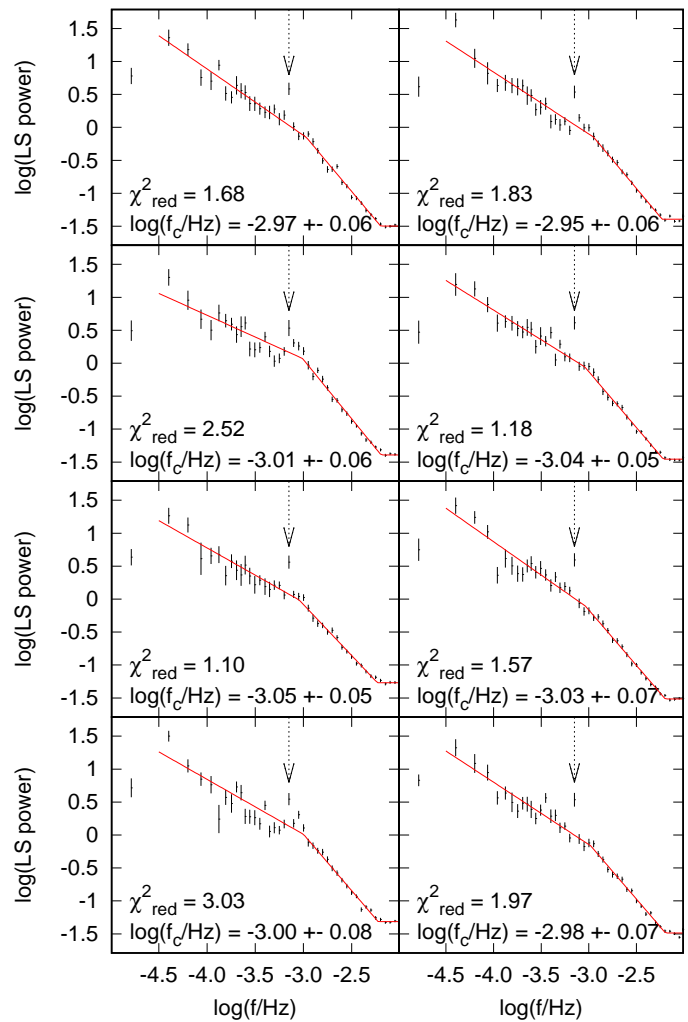}}
\caption{PDSs of V4743\,Sgr with three component fits (red lines). Eight cases are shown, because the light curve is divided into eight subsamples with duration of 10 days. Each PDS point is displayed as a vertical line showing the uncertainty interval. The latter represents the standard error of the mean. The labels report the reduced $\chi^2$ indicating the goodness of the fits, and the characteristic break frequencies $f_{\rm c}$. The vertical dashed arrows show the frequency related to the WD spin period, which was excluded from the fitting process. The PDS point with the lowest frequency is not included in the fit, because the considerably lower power is a result of detrending.}
\label{pds_v4743sgr}
\end{figure}

Dividing the V4743\,Sgr light curve into 10 day long subsamples yielded eight PDSs depicted in Fig.~\ref{pds_v4743sgr}. All PDSs show the same shape with red noise up to log($f$/Hz) $\simeq$ -2.2. Higher frequencies are dominated by the Poisson noise. However, a change of the red noise slope is obvious in all cases near log($f$/Hz) $\simeq$ -3.0. This region shows also one bin with considerably higher power which represents the optical signal related to the WD spin detected by \citet{kang2006}.

In order to search for the characteristic break frequency $f_{\rm c}$, we fitted the PDSs with a broken power law model consisting of two linear functions ${\rm log}(p) = a\,{\rm log}(f) + b$ and a constant ${\rm log}(p)$. The linear functions are connected at $f_{\rm c}$ and the constant represents the Poisson noise at the highest frequencies. The optical signal related to the WD spin, and the lowest frequency PDS point affected by the detrending were excluded from the fitting process. The resulting $f_{\rm c}$ values are summarized in Fig.~\ref{pds_v4743sgr}. The weighted mean of all eight values is $-3.01 \pm 0.02$.
%
%

\section{Discussion}

\subsection{The significance of the new measurement}

In this paper we studied the optical fast variability of a CV system, V4743\,Sgr that is likely to be in a high state and to have a hot ionized disc. We searched for characteristic frequencies in the PDSs and found a value of log($f$/Hz) $\simeq$ -3.01.

\begin{figure}
\resizebox{\hsize}{!}{\includegraphics[angle=-90]{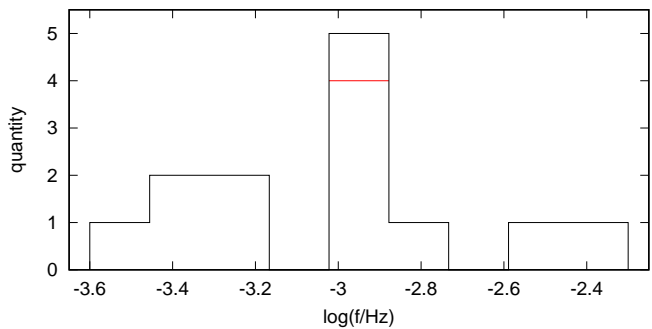}}
\caption{Histogram of the measured break/characteristic frequencies in the high state of CVs (from \citealt{dobrotka2020}, after including V4743 Sgr. The red line represents the original version.}
\label{pds_hist}
\end{figure}

\citet{dobrotka2020} summarized all the detected characteristic frequencies in CVs in optical and X-rays. They collected 12 values, and the resulting histogram (Fig.~\ref{pds_hist}) shows discrete values with dominant concentration of frequencies around log($f$/Hz) $\simeq$ -3. However, a number of 12 is too low for a significant conclusion. In this work, we increased the number of detections to 13. The histogram feature at log($f$/Hz) $\simeq$ -3 now shows 5 values compared to 4 in the histogram of \citet{dobrotka2020}.

In order to test the probability of the histogram shape being due to a random process, we performed a simulation test. We randomly selected 12 values using a uniform distribution, and we calculated the corresponding histogram. After one million simulations we found that 31.4\% of the simulations produced a peak at value 4 or higher. Therefore, the probability that CVs show a typical characteristic frequency at log($f$/Hz) $\simeq$ -3 before this study was only 68.6\%. Including the V4743\,Sgr case studied in this paper, the total number of detected frequencies increased to 13 with a maximum number of systems in the histogram of 5, at log($f$/Hz) $\simeq$ -3. This increased the confidence of the statement to 90.9\%. 

With only one additional measurement of frequency we thus transformed an uncertain idea into a much more statistically significant result.

\subsection{Origin of the flickering in V4743\,Sgr}

The root cause of the variability in  V4743\,Sgr is not straightforward to understand. The source of X-ray variability with log($f$/Hz) $\simeq$ -3 in the well studied system MV\,Lyr can be either the hot X-ray corona, or the boundary layer between the corona and the WD, and it is hard to distinguish between the two cases. However, V4743\,Sgr is an IP, which means that it does not have a boundary layer. The common structure in both systems is the geometrically thick corona surrounding the geometrically thin disc.

Magnetic CVs have truncated discs, and the inner disc edge generates the time scales of the variability. The observed $f_{\rm c}$ in IPs are very close to log($f$/Hz) $\simeq$ -3 or higher (\citealt{revnivtsev2010}, \citealt{semena2014}). Usually, $f_{\rm c}$ in the corresponding PDSs are close to the spin frequency of the WD ($P_{\rm spin}$) yielding $f_{\rm c} \times P_{\rm spin} \simeq 1$. This is observed in our PDS of V4743\,Sgr with $f_{\rm c} \times P_{\rm spin} \simeq 1.3$. The same behaviour is seen also in X-ray pulsars, with a neutron star as a central compact object (\citealt{revnivtsev2009}). The relation between $f_{\rm c}$ and $P_{\rm spin}$ is explained by corotating central objects and inner disc edges. The latter has Keplerian frequency as $f_{\rm c}$.

However, V4743\,Sgr is a post nova system, so $\dot{m}_{\rm acc}$ may be higher compared to standard IPs. The characteristic frequencies during higher $\dot{m}_{\rm acc}$ episodes shows $f_{\rm c} \times P_{\rm spin} > 10$ like in the case of X-ray pulsar A0535+26 (\citealt{revnivtsev2010}). The IP V1223\,Sgr also shows $f_{\rm c} \times P_{\rm spin} > 10$ (\citealt{revnivtsev2010}). This means that the characteristic (Keplerian) frequency at the truncation radius is considerably higher than the Keplerian frequency at the corotating radius. The same is suggested for V4743\,Sgr if we assume that it has higher $\dot{m}_{\rm acc}$ (Section~\ref{v4743sgr}). For this particular system the condition $f_{\rm c} \times P_{\rm spin} > 10$ yields considerably higher $f_{\rm c}$ than log($f$/Hz) $\simeq$ -3, and the detected $f_{\rm c}$ must be explained by another mechanism.

Therefore, we conclude that the detected log($f_{\rm c}$/Hz) $\simeq$ -3 in V4743\,Sgr is generated at the inner disc edge like in IPs, or by the central disc region with geometrically thick X-ray corona like in MV\,Lyr.

\subsection{V4743\,Sgr long-term variability}

The light curve of this quiescent nova shown in Fig.~\ref{lc_v4743sgr} exhibits clear variability on a time scale of about one day. Similar variability has been observed in VY\,Scl systems when the disc is in the high state, specifically in V794\,Aql (\citealt{honeycutt1994}, \citealt{honeycutt1998}) and FY\,Per (\citealt{honeycutt2001}). In V794 Aql, either the flares are due to disc instabilities when mass transfer from the secondary temporarily ceases (\citealt{honeycutt1994}), or, more likely, the disc remains in the high state while the variability is due to mass transfer variations from the secondary (\citealt{honeycutt1998}). Such variability on time scale of a day is never observed in known DN systems.

We also observed no DN outbursts during the 83 days of the \kepler\ observation. Standard IPs also do not show DN outbursts, but following disc instability model these are stabilised in the low cold state by the WD magnetic field and low $\dot{m}_{\rm acc}$ (\citealt{hameury2017}). Contrary, V4743\,Sgr has characteristics that are much more typical of CVs accreting at high $\dot{m}_{\rm acc}$ than of standard IPs or DN systems.

\section{Summary and conclusions}

We analysed the \kepler\ observation of the post-nova V4743\,Sgr, which is very likely to be accreting in the high state. We studied the fast stochastic variability (flickering) and found a characteristic frequency near log($f$/Hz) $\simeq$ -3 in the corresponding PDSs. We thus increase the number of detections of this characteristic frequency, and obtained further evidence that all CVs in the optical high state show this flickering frequency, as recently proposed by \citet{dobrotka2020}. The confidence of this statement was 69\% before the new data analysis contained in this paper. It increased now to 91\%, making this phenomenological evidence much more interesting and worth of further studies.

Finally, we found that the intermediate polar V4743\,Sgr is probably accreting at a higher mass accretion rate than standard intermediate polars. The accretion disc of the former should be hot with ionised hydrogen, while the latter have cold discs with neutral hydrogen (\citealt{hameury2017}). Therefore, V4743\,Sgr is similar to nova-like systems, and the source of the flickering has a different origin compared to standard intermediate polars (\citealt{revnivtsev2010}). This would be one of those rare cases in which the flickering originates in the central portion of the disk with a hot X-ray corona (\citealt{scaringi2014}) like in MV Lyr (\citealt{scaringi2014}, \citealt{dobrotka2020}).

\section*{Acknowledgement}

AD was supported by the Slovak grant VEGA 1/0408/20, and by the Operational Programme Research and Innovation for the project : Scientific and Research Centre of Excellence SlovakION for Material and Interdisciplinary Research“, code of the project ITMS2014+ : 313011W085 co-financed by the European Regional Development Fund.

\bibliographystyle{aa}
\bibliography{mybib}

\label{lastpage}

\end{document}